\documentclass[12pt,a4paper]{article}
\usepackage{amsmath}
\usepackage{appendix}
\usepackage{amssymb}
\usepackage{amsfonts}
\usepackage{graphicx}
\usepackage{lscape}
\usepackage{rotating}
\usepackage{verbatim}
\usepackage{here}
\usepackage{bm}
\usepackage{natbib}
\renewcommand{\baselinestretch}{1.55}

\newcommand{\mn}{mixture of normals}

\newcommand{\mamn}{marginally adjusted mixture of normals}

\newcommand{\mnc}{mixture of normals copula}

\newcommand{\nc}{normal copula}

\newcommand{\cdf}{cdf}

\newcommand{\KL}{Kullback-Liebler}
\newcommand{\phat}{\widehat p}
\newcommand{\Lhat}{\widehat L}
\newcommand{\KLhat}{\widehat {KL}}
\newcommand{\DGP}{data generating process}
\newcommand{\DGPS}{data generating processes}
\newcommand{\MCMC}{Markov chain Monte Carlo}

\newcommand{\veps}{\varepsilon}

\begin{document}
\title{Flexible Multivariate Density Estimation with Marginal Adaptation}
\author{Paolo Giordani \\
\textit{\ }{\small Research Division}\\
{\small Swedish Central Bank} \and Xiuyan Mun \\
{\small Australian School of Business}\\
{\small University of New South Wales}\and
Robert Kohn  \\{\small Australian School of Business}\\
{\small University of New South Wales}}
\date{}
\maketitle

\begin{abstract}
Our article addresses the problem of
flexibly estimating a multivariate density while also attempting to estimate
its marginals correctly. We do so by proposing two new estimators that try
to capture the best features of mixture of normals and copula estimators
while avoiding some of their weaknesses. The first estimator we propose is a
\mnc{}  model that is a flexible alternative to parametric
copula models such as the normal and $t$ copula. The second is a marginally
adapted mixture of normals estimator that improves on the standard mixture
of normals by using information contained in univariate estimates of the
marginal densities. We show empirically that copula based approaches can
behave much better or much worse than estimators based on mixture of
normals depending on the properties of the data. We provide fast and
reliable implementations of the estimators and illustrate the methodology on
simulated and real data.


\end{abstract}


\section{Introduction}
\label{S:introduction}

Our article is concerned with flexible and practical estimation of
multivariate densities, that is, with constructing estimators that are
computationally reliable and statistically efficient when the data
generating process is unknown. Since a multivariate density is determined by
the densities of all linear combinations of its marginal variables (that is,
by its characteristic function), this suggests that an effective
multivariate density estimator is one that can estimate reliably the
densities of all such linear combinations, and in particular the marginal
densities.

A common approach to multivariate density estimation is to use a parametric
density such as a normal or a $t$. Such densities are relatively easy to estimate
and there is extensive finite sample inference available for them; e.g.
\cite{anderson2003}.
Estimation methods and inference for more general parametric densities such
as symmetric and skew symmetric elliptic densities are also available; e.g.
\cite{genton2004}.
An important advantage of parametric densities is that they
can be applied to high dimensional problems because of the relatively small
number of parameters involved. However, the small number of parameters can
also be a major disadvantage if the data generating process differs
significantly from the parametric model. For example, if bivariate data
is modeled as a bivariate normal distribution then just one parameter (the
correlation coefficient) is available to capture the dependence between the
two marginals.

Parametric or semiparametric copula models such as the multivariate normal
or $t$ \citep{joe1997,nelsen1999}
provide extra flexibility in density
estimation by separately modeling the marginal densities and then linking
them through a joint parametric model of dependence such as the multivariate
normal or $t$ distribution \citep{joe1997,nelsen1999}.
For example, a
Gaussian copula approach to density estimation models each marginal
separately using a parametric (or possibly nonparametric) model and then
transforms each marginal to a standard normal. These transformed marginals
are then modeled as a multivariate normal distribution. See also
\cite{demartamcneil2005} for two parametric generalizations of the $t$ copula, the
\textit{skewed} $t$ and the \textit{grouped} $t$ copulas. We note that an
attractive feature of copula models is that the marginal densities implied
by the copula density are the \emph{same} as those originally proposed for
the marginals. In addition, by transforming each marginal to a standard
distribution such as a standard normal there is some hope that the joint
distribution of the transformed variables will also behave \lq nicely\rq.

In general, however, the use of parametric copula models to capture
dependence is likely to have the same advantages and disadvantages as
conventional parametric models. Thus, suppose that we have bivariate
observations with each marginal having support on $m$ points so that the
joint distribution has support on $m^{2}$ points. If we model the bivariate
distribution using a Gaussian copula then we may hope to capture the
distribution of each marginal that has support on $m$ points through an
appropriate model for that marginal. However, the Gaussian copula only
allows one extra parameter (the correlation coefficient) to capture the
joint dependence contained in the remaining $m^{2}-2m$ points of support.
Clearly, such modeling problems increase as $m$ increases and as the
dimension $p$ of the multivariate vector increases.

To overcome the problems encountered using parametric methods, there is a
large and growing literature on nonparametric density estimation. One
popular approach is kernel density estimation. \cite{sheather2004}
surveys the
univariate case and makes it clear that the critical aspect of the method is
the choice of smoothing or bandwidth parameter. We infer from Sheather's
article that it will be challenging to successfully apply kernel density
estimation in higher dimensions. A second approach is to use finite mixture
of normals; see for example \cite{McLachlan2000}
for a discussion of univariate and multivariate
approaches to estimation, and \cite{RichardsonGreen97} and
\cite{roederwasserman97}
for fully Bayesian univariate analyses. A third approach is
to use Dirichlet Process Mixtures (DPM) which in many applications is
equivalent to an infinite mixture of normals with constraints on the mixing
probabilities; see \cite{greenrichardson2001} for a comparison of finite
mixture models and DPM. We note, however, that one of the problems
encountered by estimating densities by
mixture models, especially in higher dimensions, is that in
finite samples the implied model for each marginal may not be even close to
the best model for that marginal. To understand why, consider the
example of a multivariate distribution with $p$ independent
marginals (possibly all identical)
where each marginal is a mixture of $m$
normals. Then the joint distribution is also a mixture of normals,
but with $m^{p}$ components so that for moderate values of
$m$ and $p$ it will
usually be unrealistic to fit a multivariate mixture model with so
many components as such a model will be
highly over-parametrized. See Section~\ref{SSS:direct vs indirect} for further
discussion.

We propose two new estimators of multivariate densities that attempt to
capture the advantages of parametric copula estimators and mixture of
normals estimators while being more robust to their weaknesses. The first
estimator is a \emph{\mnc{}} (MNC) where we flexibly
estimate each of the marginals, transform the marginals to standard normal
distributions, and then flexibly estimate the joint distribution through a
mixture of normals with approximately normal marginals. This defines a
more flexible copula than currently available in the literature.

We show through simulation that the mixture of normals copula performs well
when the data are generated by a normal or $t$ copula, but that a normal and
$t $ copula can provide a poor fit when the assumed model is incorrect. A
mixture of normals and a mixture of normals copula are both universal
approximations to any multivariate density when the number of components is
estimated from the data. That is, given enough data they will provide
accurate approximations to any multivariate density. However, and contrary
to our initial intuition, either can greatly outperform the other in any
given dataset. In moderate to large dimensions, copulas
tend to perform poorly when the joint distribution is non-normal but
well modeled by a
mixture of normals with few components. Conversely, a mixture of normals can
be grossly inadequate when a normal or $t$ copula fits the data
well. An example of this occurs when $p$ variables
have non-normal distributions but weak dependence.

Motivated by these findings we introduce a second density estimator,
the \emph{marginally adapted mixture of normals}, where we correct the
mixture of normals density estimator by a factor that reflects the
difference between the univariate and multivariate estimates of the marginal
distribution of each variable. If the marginals are well approximated\ by
univariate estimators, then  this estimator can be shown to be at least as
good as a mixture of normals in a sense made precise in
Section \ref{S:marginally adapted densities}.
In practice, the marginally adapted mixture
of normals can be expected to improve on the standard mixture of normals
whenever the density cannot be well approximated using only a small number
of components. We also note that marginal adaptation of
multivariate densities is a general concept that can be applied outside the
mixture of normals framework.

The mixture of normals model is the basic building block of our estimators.
This is not a simple model to fit because the likelihood can be multimodal
and badly behaved. See for example \citet[][Chapter 3]{McLachlan2000}.
We implement our methods using the computationally efficient and
fast stochastic approximation algorithms described in Appendix~\ref{B:stoch_approx}.

\section{Estimating the normal, $t$ and mixture of normals copulas\label%
{S:copula estimation}}

We briefly introduce copulas and refer to \cite{joe1997}
and \cite{nelsen1999} for a modern treatment. The multivariate function
${C}(\bm{u})= {C} (u_{1},...,u_{p})$ is a copula if it is a multivariate cumulative distribution
function (cdf) on the $p$-dimensional cube $[0,1]^{p}$ with uniformly distributed
marginals. If $U_{1},\dots ,U_{p}$
are the corresponding $p$ marginal variables then
\begin{equation}
{C}(\bm{u})=\Pr(U_{1}\leq u_{1},...,U_{p}\leq u_{p}).  \label{e:copula}
\end{equation}%
We can construct a copula model explicitly for a random vector $%
\bm{y} = (y_1, \dots, y_p)$ by choosing a copula
${C}(\bm{u})$ with $\bm{u}$ obtained by transforming
each $y_{j}$ to a uniform using its marginal \cdf{}
so that $u_{j}=H_{j}(y_{j})$. However, the more popular
copulas are defined implicitly on transformations of $\bm{u}$ as follows. Suppose
that $\bm{x}$ is a multivariate $p$ dimensional random variable with density $%
f(\bm{x}) $ and \cdf{} $F(\bm{x})$, and corresponding marginal densities and cdf{}'s $%
f_{j}(x_{j})$ and $F_{j}(x_{j}),j=1,\dots ,p$. We now define one to one
transformations of $\bm{x}$ to $\bm{y}$ by
$F_{j}(x_{j})=u_{j}=H_{j}(y_{j})\ ,j=1,\dots ,p$\ .
Then the implied density of $\bm{y}$ is
\begin{equation}
p(\bm{y})=f(\bm{x})\prod\limits_{j=1}^{p}\biggl  (  h_{j}(y_{j})/f_{j}(x_{j})\biggr ) .  \label{eq: Py}
\end{equation}%
For example, a normal or Gaussian copula is defined implicitly by taking $x$
as multivariate normal with zero mean and with standard normal marginals.

Joint estimation of both the copula and marginal parameters is challenging
even in problems of moderate dimension. The standard approach
fits separate models to each marginal and treats the resulting
distributions as fixed when estimating the copula. We also follow this
two-stage approach. \cite{joe1997} and in the Bayesian literature \cite{pitt2006}
fit parametric distributions to the marginals. A more common
procedure estimates the marginals nonparametrically using the marginal
empirical distribution functions, as in \cite{demartamcneil2005}. We model
the marginals as a mixture of normals. This is more computationally demanding
than the nonparametric approach based on the marginal empirical
distributions, but should give a more efficient estimate of the joint
distribution function in small samples and is better suited to incorporating
regression effects.

Appendices \ref{SS:Normal Copula} and \ref{SS: t copula} outline how we estimate the normal and $t$ copulas.

\subsection{Mixture of normals copula}
\label{SS: MN copula}

We propose to define and estimate a \mnc{} as
follows. Steps one and two are the same as for a normal copula. In step
three, we fit a mixture of normals to $x,$ with the number of components chosen
by BIC. The parameters of the mixture of normals implicitly define the
copula.

The one component case is a normal copula, estimated exactly as in appendix %
\ref{SS:Normal Copula}. With more than one component, the parameters
estimated in step three will not imply exactly standard normal marginals.
This discrepancy between steps two and three implies a small efficiency loss
in small samples, but poses no theoretical problem, since any multivariate
distribution for $\bm{x}$ implicitly defines a copula as long as the density $p(\bm{y})$
is computed as in equation \eqref{eq: Py}. When evaluating or drawing from $%
p(\bm{y})$ one must therefore take into account that the marginal distribution $%
f_{j}(x_{j})$ in this case is not standard normal but a mixture of normals.
Moreover, the one-to-one transformation $x_{j}=F_{j}^{-1}(H_{j}(y_{j}))$
which implicitly defines the copula requires that for use in \eqref{eq: Py}
we recompute $\bm{x}$ using the mixture of normals parameters to define $F_{j}$.

\section{A comparison of estimators \label{S:comparison}}

This section investigates empirically the performance of various copula based estimators,  a mixture of normals estimator and a skew $t$ estimator
for different \DGP{} (DGP). The
main conclusions are that (i)~we can expect the mixture of normals copula to
rarely perform much worse than a $t$ copula in problems of moderate
dimensions, while the contrary need not hold; (ii)~both $t$ and mixture
of normals{} copulas can fit either much better or much worse than mixture
of normals{} depending on the \DGP.

\noindent
\subparagraph{Simulation design.}

In all the simulation experiments we use a
sample of $n = 500$, $p = 5$ variables and 50 replications. All the
copula \DGPS{} share the same marginals, which are a mixture of normals.
The number of components for all mixture of normals,
whether in copulas or stand-alone,
is chosen by BIC in the range $1$ to $10$ (where $10$ was never chosen in our
simulations). The estimation process and the tuning parameters for the mixture of
normals estimators are
described in Appendix~\ref{B:stoch_approx}.

We use the \KL{} divergence and the $L_2$ distance of the estimate from
the true model to compare the performance of the various estimators.
The results are reported relative to a given estimator, usually
the estimator corresponding to the \DGP{}. The
\KL{}
divergence between the estimate $\widehat{p}(\bm{y})$ and the true density $p(\bm{y})$ is
\begin{align} \label{e:KL div}
KL(p,\phat) & = \int p(\bm{y})\log \biggl (\frac{p(\bm{y})}{\phat(\bm{y})}\biggr )d\bm{y}\ .
\end{align}%
We estimate \eqref{e:KL div} by
\begin{align} \label{e:KL div est}
\KLhat (p,\phat) & = N^{-1}\sum_{i=1}^{N}\log \biggl (\frac{p(\bm{y}_i)}{\phat (\bm{y}_i)}\biggr )\ ,
\end{align}%
where $\bm{y}_{i},i=1,\dots ,N$ are a sample of $N=5000$ observations drawn
from the $p(\bm{y})$ and which are different from the observations used to estimate
each model. The $L_2$ loss is defined as
\begin{align*}
L_2(p,\phat) & = \int \biggl (  p(\bm{y}) - \phat (\bm{y}) \biggr )^2 d\bm{y}
\end{align*}
and is estimated by
\begin{align*}
{\Lhat}_2(p,\phat) & = N^{-1} \sum_{i=1}^N  \biggl ( p(\bm{y}_i) - \phat (\bm{y}_i) \biggr )^2 /p(\bm{y}_i) \ ,
\end{align*}
where the $\bm{y}_i$ are defined similarly to \eqref{e:KL div est}.

Our simulations considered a number of \DGPS{} which are discussed below and the
following estimators. (a)~The \nc{} (NC).
(b)~The $t$ copula with estimated degrees of
freedom (tC). (c)~The mixture of normals copula (MNC). (d) The Clayton, Frank and
Gumbel Archimedian copulas described briefly in Appendix~\ref{A: copulas}.   (e)~The mixture of normals estimator (MN).
(d)~The multivariate skew $t$ estimator (ST) in \cite{Sahu03}, whose aim is to
capture both multivariate skewness and kurtosis. This estimator is described
briefly Appendix~\ref{A: multivar skew t}.
For all copulas, the marginals were estimated by a mixture of normals.

A more comprehensive set of simulations is reported in \cite{giordani_mun_kohn_08}, which is an extended version of the current article.

\subsection{Normal copula \DGP} \label{SS: nc dgp}
The \DGP{} for this simulation has
marginals that are mixtures of normals with three components, with means $\mu _{j,1}=0,$\
$\mu _{j,2}=-3,$\ $%
\mu _{j,3}=3$, component probabilities $\pi _{j,1}=0.6$ and $\pi
_{j,2}=0.2, $ and component variances $\sigma _{j,1}^{2}=1,$\ $\sigma
_{j,2}^{2}=9,$\ $\sigma _{j,3}^{2}=0.1,$\ for $j=1,...,p,$. The copula is  $N(0,\bm{V})$ with $\bm{V}=0.5\bm{I}+0.5%
\bm{i}\bm{i}^{T},$\ and $\bm{i}$\ the unit vector.
Table~\ref{T:NC}
reports the median of the logarithm of the ratio of KL divergence for a particular
estimator and the KL divergence of the $t$ copula estimator over the 50
replicates. If we multiply each entry in the table by 100 then we can interpret
each entry as approximately the median percentage increase in KL divergence of the
particular estimator relative to the $t$ copula estimator.
The table also shows
those log ratios that are {\em not} significantly different from 0
at the 1\% and 5\% levels as judged by the Wilcoxon rank sum test. The entries for
the $L_2$ loss function are interpreted similarly. We work with the logarithm of
the ratios of the loss functions as these are distributed closer to normality than
the ratios them selves. When two estimators perform similarly relative to a loss
function, we would expect the median of the logarithm of the ratios to be approximately zero and this is why we report this median.

Since the number of components in the mixture of normals{} copula is chosen
by BIC rather than fixed, we expect it to perform nearly as well as a normal
copula even when the latter generates the data. This is
confirmed by the simulation results reported in Table~\ref{T:NC}.
The normal, $t$\ and mixture of normals{} copulas perform similarly.
In particular, the BIC criterion
almost always selects one component for the \mnc{} so the
loss of efficiency from estimating a \mnc{} copula is negligible. However, the losses
for the three Archimedian copula are substantial despite the marginals being estimated flexibly. The losses for the mixture of normals estimator and the skew t estimator can also be substantial.

\subsection{Archimedian copula \DGPS} \label{SS: archimedian dgp}
Table~\ref{T:Clayton} presents simulation results when the \DGP{} is a Clayton
copula with parameter $\theta = 5$ and with the same marginals as in Section~\ref{SS: nc dgp}. The table shows that the \mnc{} performs best overall,  and that
the two other Archimedian copulas do not perform very well when the true
\DGP{} is a Clayton copula.

\subsection{Comparing a mixture of normals copula to a mixture of normals} \label{SS: MNC vs MN}
One may expect a mixture of normals{} copula and a mixture of normals{} to
perform similarly in any given dataset. In fact, either can greatly
outperform the other depending on the characteristics of the density to be
approximated. It is useful to consider two reasons why a mixture of
normals{} copula may fit better (worse) than a mixture of normals: (i)
direct estimation of the marginals is more (less) accurate than indirect
estimation through the joint distribution; (ii) the transformed variables $x,$
with normal marginals, are easier (more difficult) to fit with a \mn{} than the
original variables $y$. We now discuss this issue conceptually and report
some simulation results.

\subsubsection{Direct vs indirect estimation of the marginal densities.} \label{SSS:direct vs indirect}

Indirect estimation of the marginal densities through the joint distribution
is more efficient if the model for the joint is correct, but less robust to
model misspecification. Consider a deceptively mild form of model
misspecification, namely over-parameterization. Assume that the \DGP{} is a
mixture of normals. Define the degree of over-parametrization as the number
of valid exact restrictions not imposed on the parameters of a mixture of
normals{} over the total number of estimated parameters. A mixture of
normals{} copula can be expected to outperform a \mn{} when the degree of
over-parameterization is high.

For example, suppose that $p$ independent variables are
each generated by the same univariate mixture with $m$ components. That is,
the marginal densities are all identical. However, the joint distribution is
a mixture of normals{} with $m^{p}$\ components. In this case, a mixture of
normals{} quickly becomes highly over parametrized as $p$ gets larger, while
a mixture of normals{} copula will fit the marginals parsimoniously and then
use one component for the copula. For medium and large $p$, a mixture of
normals{} copula should therefore outperform a mixture of normals{} in this
example. The simulations reported in Table~\ref{T:NC} and other simulations not reported in the article confirm this analysis. The
ability of a \mn{}  to fit the \DGP{}  deteriorates very quickly with $p.$ The
results are even worse if we set $\bm{V}=\bm{I}$ rather than $\bm{V}=0.5\bm{I}+0.5\bm{ii}%
^{T} $ (not reported).

Less extreme cases are likely to occur in empirical applications. For
example, if the variables can be divided into $l$\ groups, each a mixture of
$m$\ components independent of the other group, a \mn{}  for the joint
distribution requires $m^{l}$\ components. In these situations the
over-parametrization will typically result in poor fit and in the
model selection
criteria choosing less components than in the \DGP{} (a mixture of factor
analyzers should perform better in these cases).

\subsubsection{Fitting $\bm{x}$ vs fitting $\bm{y}$.} \label{SS: y vs x}

\begin{figure}[tbp]
\centering
\includegraphics[angle=0, width=1.0\textwidth,
height = 0.45\textheight]{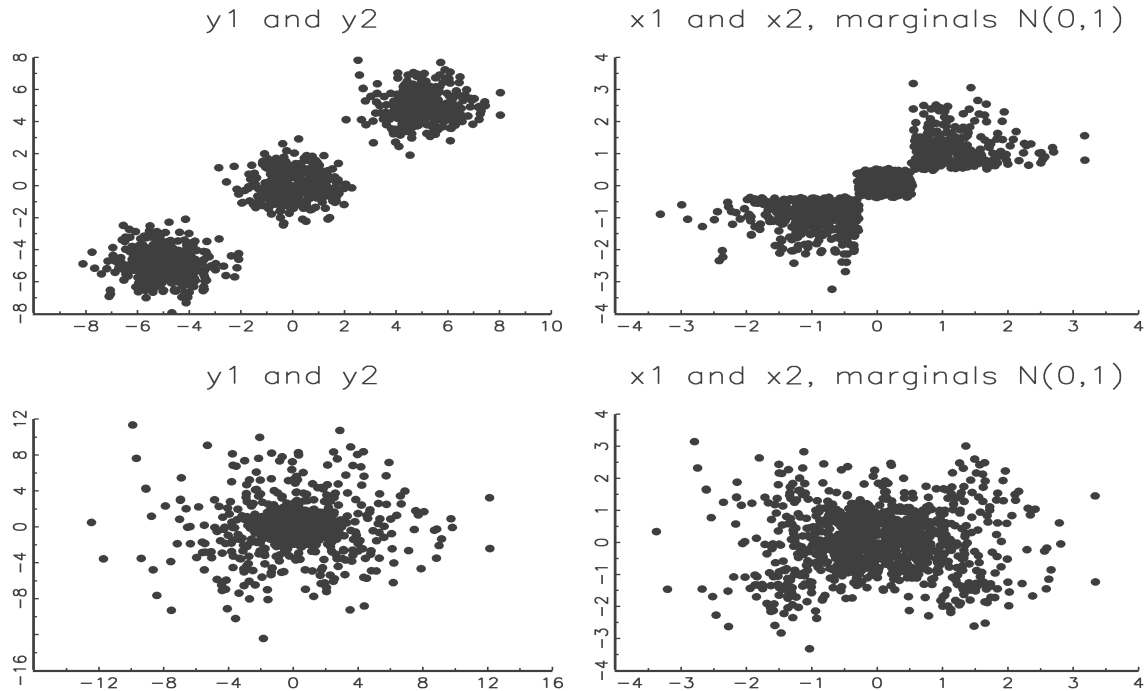}
\caption{First row. 1000 observations
generated by a bivariate mixture of 3 normals $(y_1, y_2)$\ and corresponding
$x_1, x_2$ with $N(0,1)$ marginals obtained through the true marginal densities of
$y_1$ and $y_2$. Second row. As for the first row, observations generated by a
scale mixture of two normals. See Section~\ref{SS: y vs x}. }
\label{F:Figure1}
\end{figure}

In the simulations summarized in Table~\ref{T:NC},
$\bm{x}$ is multivariate normal and $%
t$ respectively, making it easier to model than $\bm{y}$.
 However, when the data
cluster, $\bm{x}$ can be much more difficult to fit than $\bm{y}$. The cluster
representation evident in $\bm{y}$ can be severely distorted in the $\bm{x}$, making
the multivariate distribution of $\bm{x}$ extremely complex.
Consider data generated by a mixture of three well-separated bivariate
normals%
\begin{equation*}
p(\bm{y})=\frac{1}{3}\phi _{2}(\bm{y};\bm{0},\bm{I})+\frac{1}{3}\phi _{2}(\bm{y};-5\bm{i},\bm{I})+%
\frac{1}{3}\phi _{2}(\bm{y};5\bm{i},\bm{I}),
\end{equation*}%
where $\bm{i}$ is defined above and $\phi _{p}(\bm{\mu} ,\bm{\Sigma} )$ is a $p$%
-dimensional multivariate normal density with mean $\bm{\mu }$ and variance $%
\bm{\Sigma} $. The first row of Figure~\ref{F:Figure1} shows 1000 observations
generated from this \DGP{} together with the $\bm{x}$ obtained through the true
marginal densities. It is clear that the fact that  $\bm{x}$ has standard
normal marginals is of little comfort, as the joint distribution of $\bm{x}$ is
extremely difficult to model. Overlapping clusters also cause trouble for
copulas, though not as dramatically. Consider data generated by a scale mixture
of two normals
\begin{align*}
p(\bm{y})=0.6\phi _{2}(\bm{y};\bm{0},\bm{I})+0.4\phi _{2}(\bm{y};\bm{0},16\bm{I}),
\end{align*}%
from which we generate 1000 observations, as displayed in the second row of
Figure \ref{F:Figure1}. Clearly more than two components are needed to capture
the joint density of $\bm{x}$ adequately.

The simulation results reported in Table~\ref{T:MN} confirm this analysis. To
emphasise the point that the clusters need not be separated for copulas to
work poorly, the data are generated by a scale mixture of two normals
\begin{equation*}
p(\bm{y})=0.7\phi_p (\bm{y}; \bm{0},\bm{I})+0.3\phi_p (\bm{y}; \bm{0},4\bm{I}+5\bm{ii}^{T}).
\end{equation*}%
The normal copula performs very poorly. The $t$ copula is better but still
poor. The \mnc{}  improves on the $t$ copula but still
produces large losses compared to a \mn{}.

\section{Marginally adapted multivariate densities
\label{S:marginally adapted densities}}

The discussion in the previous section highlights an important trade-off
involved in estimating multivariate distributions. To capture the dependence
structure of a set of variables parsimoniously we usually need to place
strong constraints on their marginal densities and, conversely, focusing on
the marginal densities may make it harder to model the dependence
effectively. Motivated by these results we introduce the class of \textit{%
marginally adapted densities}. The idea is to fit a multivariate density to
the original data and then correct it by a factor reflecting the discrepancy
between the marginal distributions implied by the multivariate model and
those fitted directly to each variables. If the second set of marginals is
more accurate than the first, the marginally adapted density is likely to be
closer to the true density in a sense that is made precise below.

Suppose that $f(\bm{y})$  and $h(\bm{y})$ are $p$  dimensional densities
with respect to Lebesgue measure with marginal densities $f_i (y_i ), h_i(y_i)$ ,
$i = 1, \ldots ,p$.
For $0 \leq \veps\leq 1$, let $f_{i, \veps} (y_i) = (1-\veps) f_i(y_i) + \veps
 h_i(y_i)$, for $i=1, \dots, p$. Then each $f_{i, \veps} (y_i)$ is a density and
  $h_i(y_i)/f_{i, \veps} (y_i)\leq 1/\veps$ for $\veps> 0$. Let
\begin{align} \label{e: marg adapt}
p_{h,\veps} (\bm{y}) &  = k_{h,\veps} f(\bm{y}) \prod\limits_{i = 1}^p
h_i (y_i )/f_{i,\veps} (y_i )\ ,
\end{align}
where $k_{h,\veps} $
is a normalizing constant that makes $p_{h,\veps} (y)$
a density for $\veps>0$.
We say that $p_{h,\veps} (\bm{y})$ is
the density of $f$ adjusted for the marginals $h_i $.
The following result will give necessary and sufficient conditions for $p_{h,\veps} (\bm{y})$ to be closer to $h$ than $f$ in KL divergence,
where the KL divergence is defined by \eqref{e:KL div}.

\noindent
{\em Lemma 1}. Suppose that $f$
and $h$ are $p$ dimensional multivariate densities with the marginals $f_i $
and $h_i $. Then, for $0<\veps\leq  1$,
\begin{align*}
KL(h,f ) -  KL(h,p_{h,\veps} ) & = \log (k_{h,\veps} )+
 \sum\limits_{i = 1}^p {KL(h_i ,f_{i,\veps} )}  \ .
\end{align*}
\noindent
{\em Proof}. \\
\begin{align*}
 KL(h,f ) - KL(h, p_{h,\veps}) &
 = \int {h\log \left( {p_{h,\veps}(\bm{y}) /f(\bm{y}) } \right)} d\bm{y} \\
 & = \log (k_{h,\veps} ) + \sum_{i=1}^p \int  h(\bm{y}) \log \left( h_i(y_i) / f_{i,\veps}(y_i)\right) d \bm{y} \\
 & = \log (k_{h,\veps} ) + \sum_{i=1}^p \int  h(y_i) \log \left( h_i(y_i) / f_{i,\veps}(y_i)\right) d y_i
 \end{align*}
 and the result follows.

 The lemma shows that $KL(h,f ) >  KL(h,p_{h,\veps} )$ if
 \begin{align} \label{e: domin}
 \log (k_{h,\veps} )+
 \sum\limits_{i = 1}^p {KL(h_i ,f_{i,\veps} )} & > 0 .
 \end{align}
We note that the sum $\sum_i {KL(h_i ,f_{i,\veps} )} > 0$ unless $h_i = f_{i,\veps}$ almost everywhere for all $i$ by the properties of the KL divergence. Thus, $KL(h,f ) >  KL(h,p_{h,\veps} )$ is likely to hold if $k_{h,\veps}$ is close to 1. We also note that the condition \eqref{e: domin} can be verified for any given $f$ if we know the marginals $h_i$, but not necessarily the joint  density $h$.

We apply Lemma~1 as follows. Let $h$ be the true multivariate density and $f$
an approximation (or estimate) of it. Suppose that we know the marginals
of $h$, but not $h$ itself. Then we can compute $k_{h,\veps}$ and determine whether $KL(h,f ) >  KL(h,p_{h,\veps} )$ using
\eqref{e: domin}. We note the following:
\begin{enumerate}
\item
Condition \eqref{e: domin} can be verified for any given $f$ if we know the marginals $h_i$, but not necessarily the joint  density $h$.
\item
The marginally adaptive density estimator \eqref{e: marg adapt} is not in general a copula since its marginals are not necessarily $h_{i}(y_{i}).$
\item
We also note that the result of the lemma  is
very general and does not require either $f(\bm{y})$ or the $h_{i}(y_{i})$ to be
a \mn{}. It can be applied to both simpler more complex models. A simple model
could be a multivariate $t$ distribution (or any other parametric
multivariate density), with marginals also $t$ distributions (but each with
possibly different degrees of freedom) or a \mn{} or nonparametric kernel
density estimates. A more complex model for $f(\bm{y})$ could be a factor model
in high dimensional data.
\item
Lemma~1 assumes that the marginal distributions $h_{i}(y_{i})$ are
known. In practice the marginals $h_{i}(y_{i})$ are estimated from the data
and we take $\veps = 0.05$.
\end{enumerate}

\subsection{Marginally adapted mixture of normals} \label{ss: mamn}
The \textit{marginally adapted mixture of normals} (MAMN) estimator
specifies $f(\bm{y})$ as a multivariate mixture of normals density, so that its
marginals are also \mn{}  and therefore straightforward to compute. The \mamn{}
retains the ability of the \mn{}  to model clustering (overlapping or not) while
reducing the risk of poor fitting of the marginal densities. We
choose to model $h_{i}(y_{i})$ as a univariate mixture of normals, but we
could use any other model estimate.

\subparagraph{Estimating $k_{h,\veps}$.}

In general there is no analytical expression for the normalizing constant
 $k_{h,\veps}$ so we estimate it by importance sampling as
\begin{align} \label{e: est k}
k_{h,\veps}^{-1} & \simeq \frac{1}{M}\sum_{t=1}^{M}\prod%
\limits_{i=1}^{p}\left \{ h_{i}(\bm{y}^{t})/  f_{i,\veps}(\bm{y}^{t})\right \},
\end{align}%
where $\bm{y}^{t}$ is the $t$th draw from from $f(\bm{y}).$ This estimate converges
to the true value (typically slightly lower than one)\ as $M\rightarrow
\infty $.

In practice, to prevent bad behavior of the estimates, we constrain  $%
h_{i}(\bm{y}^{t})/f_{i,\veps}(\bm{y}^{t})$ to lie between 0.02 and 50
(set to 0.02 if smaller and to 50 if larger).
We note that when the ratios of
$h_i/f_{i,\veps}$ lie
outside the bounds of $0.02$ and 50 for a considerable number of the iterates then
the resulting estimate of $k_{h,\veps}$ may be unreliable.
This information is useful
because it indicates that the marginals implied by $f$ are very poor and suggests
that marginal adaptation will be beneficial but that first the estimate $f$ needs
to be improved in order to estimate $k_{h,\veps}$. For example, adding more components to the \mn{}
has helped considerably in our experience; considering a mixture of $t$ densities
should also help because of its
fatter tails. This \lq problem\rq{} of highly variable weights is more likely to
happen in higher dimensions, which is just a way of saying that in higher
dimensions a \mn{} has more trouble capturing the marginals.

\subparagraph{Sampling from the \mamn{}.}

Drawing from a \mamn{} can be performed by independent Metropolis-Hastings
using $f(\cdot )$ as a proposal density.

\subparagraph{Simulation results using the \mn{} \DGP .}

Table~\ref{T:MN} shows that the efficiency loss of the \mamn{} estimator is small
compared to a \mn{} when a \mamn{} is estimated
on data generated by a \mn{}.

\subparagraph{Multivariate mixture with a non-Gaussian component.}

The results above suggest that the \mn{}, \mamn{} and \mnc{} estimators are general
approaches to multivariate density estimation. We now study the performance
of these three estimators when the data is generated by a finite multivariate
mixture that is not a mixture of normals. This is an important situation because a
\mn{}  estimator may not provide good estimates for the marginals while the previous
results suggest that copulas do not estimate multivariate mixtures well.

The data generating process is a mixture with four components. The first three are normal with
\begin{align*}
\bm{\mu}_1 &  =\bm{0}_{p \times 1}, \quad \bm{\mu}_2  =\left(\begin{array}{cc}
\bm{0}_{p-1 \times 1}	\\ -3
\end{array}\right), \quad \bm{\mu}_3  =\left(\begin{array}{cc}
\bm{0}_{p-1 \times 1}	\\ -6
\end{array}\right)\ ,  \\
\end{align*}
and $\bm{\Sigma}_1  =\bm{I}_p, \quad \bm{\Sigma}_2  =2\bm{I}_p, \quad \bm{\Sigma}_3  =\bm{I}_p $,
while the fourth is uniform with support in the range $-10$ to 10. The mixing
proportion for the normal components is $0.66$ which is split between the three in
the following proportions, $[0.6,0.2,0.2]*0.66$ and $\pi_4=0.34$. This example is related to the example in McLachlan and Peel~(2000, p.231).

Table~\ref{T:MNU} shows the efficiency loss of the \mn{} and the
\mnc{} estimators relative to the \mamn{} estimator.
 We also created boxplots (not shown) of these log loss ratios for all 50 replications. The table and boxplots show that the \mamn{} clearly outperforms both the \mn{} estimator and the \mnc{} for this example.

\section{Regression density estimation}
\label{S:regression density} The previous sections considered pure density
estimation without any covariates. It is usually important to allow for some regression effects as well. In our article we consider the simplest such case
$\bm{y} = \bm{W\beta} +\bm{e}$\ ,
where $\bm{W}$ is a matrix of regressors excluding the constant and the error $\bm{e}$
has an unknown density $p_{\bm{e}}(\bm{e})$.  It is straightforward to extend all our estimators for this case.

\section{Real Examples} \label{s: real examples}
\subsection{Fama and French Three-Factor Model}\label{ss: fama french}
Financial returns typically display non-Gaussian behavior. Moreover, construction of an optimal portfolio or computation of risk measures like value-at-risk require a model of the joint distribution of returns.
We now consider the well-known  \cite{fama_french1993} three-factor model used by many researchers and practitioners to model financial returns.
\begin{equation}
r_{j,t}=\beta_{M,j} r_t^M + \beta_{SMB,j} SMB_t + \beta_{HML,j} HML_t +\epsilon_{j,t} \ ,
\end{equation}
where $r_{j,t}$ is the excess return (i.e. the return minus the risk-free interest rate) of asset $j$ in period $t$,  $r_t^M$ is the market excess return, $SMB_t$ and $HML_t$ refer to the size and value factors. We use monthly data for the period 1968m1-2007m12 for 5 industry portfolios: (1) Consumer; (2) Manufacturing; (3) High Tech; (4) Health; (5) Other. The data are taken from Kenneth French's website (http://mba.tuck.dartmouth.edu/pages/faculty/\\ken.french/data-library.html), to which we refer for details.

We used ten-fold cross validation (on reshuffled data, so the test samples have no temporal dimension) to rank the models with the results reported in Table $\ref{TableIndPF}$. The table shows that: (a) the t copula and the
skew t-distribution both have small degree of freedom; (2) both \mn{} and \mnc{} choose two components for all ten subsets;  and (3) the \mamn{} estimator
performs the best. Most of the empirical work on the Fama and French three-factor model assumes that the errors are normally distributed or t-distributed.  Our  results show that both are insufficient representations of the distribution of the errors and that it is important to model the marginals well to improve prediction.

\subsection{Realized volatility of bonds and stocks}  \label{ss: realized volatility}
A major advance in modeling volatility in finance over the past decade has been the
construction of volatility estimators  that are constructed using intraday returns.  The treatment of volatility as observed rather than latent has enabled model-free analysis of its distributional and dynamic properties. A number of authors including 
\cite{Anderson01a} and \cite{Thomakos03}
found realized volatilities exhibit long-term memory and are right-skewed and leptokurtic while the logarithm realized volatilities display approximate Gaussianity. Beside statistical studies on realized volatility, the economic benefits of realized volatility have also been documented by \cite{Fleming03} who reported that investors are willing to pay more to capture the performance gains in a  volatility-timing strategy implemented using realized volatility estimated with intraday returns relative to daily returns.

We model the logarithm of daily realized volatility of S$\&$P 500 and US bond futures over a period of 10 years from 1997m1 - 2006m12. The daily realized volatilities are computed by summing the squared intraday returns over 5-minute intervals for each day.  The bivariate realized volatilities are modeled as a vector autoregressive model with 20 lags assuming several
different distributions for the errors. The results are reported in Table~$\ref{TableRV}$. Under a mixture of normal specification, more than two components are needed to capture the distribution of the errors. The number of components for the marginal estimation for the cross-validated subsample is two for S$\&$P 500 futures and four for the bond futures while the number of components for the joint distribution is two. This could explain why \mn{} performs relatively poorly and the benefits of separately modeling marginals are apparent with the \mamn{} being the  best
followed by the copula models.

\subsection{Gene expression data} \label{ee: gene expression}
Malaria is an infectious diseases caused by the parasitic protozoan genus plasmodium. It is a major concern in developing countries. The study of plasmodium molecular biology is thus of great importance in order to develop effective anti malaria treatment and vaccine strategy.  In this example, we consider the relative expression level of 4221 parasite genes taken at 46 time points over a 48 hour period of the life cycle of the parasite. The gene expression data is further processed by \cite{Jasra07} using K-means clustering and principal component analysis to reduce the number of observations from 4221 to 1000 and the number of variables from 46 to 6. We fit our models to this dataset and present the results in Table~\ref{TableGene}. \cite{Jasra07} model the multivariate density of the reduced data set as a mixture of multivariate $t$ densities with common degrees of freedom and use MCMC methods to estimate the model. Their results suggest that the number of components is between 2 and 7.
Here we allow the \mn{} and \mnc{} to select the number of components using the Bayesian information criterion (BIC) for up to ten components. The estimated average number of components for the joint distribution for both models is about five and none of the models required more than eight components for any cross-validation subsample. \mamn{} is the best estimator to use for this example given that the six marginal distributions need on average about 1.8 components to fit the dataset well.


\section{Conclusions\label{S: Conclusion}}
Both copula models and mixture of normals{} models provide estimators of
multivariate densities. Our article identifies deficiencies in both these
estimators and proposes flexible modifications of these estimators that
attempt to simultaneously estimate correctly the dependence structure as
well as the marginals. The major challenge is to be able to extend these
estimators to perform well in moderate and high dimensions.

\section*{Acknowledgement}
We  thank Professor Jiang for the current form of Lemma~1 and Professor Jasra for the genome dataset.


\begin{appendices}
\section{Estimating normal and t copulas}
\subsection{Normal copula\label{SS:Normal Copula}}

We estimate a normal copula as follows: (a)
Estimate each marginal as a mixture of normals, with the number of
components  chosen by BIC. (b)~Use the estimates to construct the cumulative density of each variable
and transform the original variables $\bm{y}$ into latent variables $\bm{x}$ where
each element of $\bm{x}$ is standard normal. (c)~Fit a multivariate normal distribution $N(\bm{0},\bm{V})$ to $\bm{x}$.

To maximize the statistical efficiency in step 3, the covariance matrix $\bm{V}$
should be constrained to have unit diagonal elements. According to \cite{mcneil2005}, this is very slow in high dimensions. It is therefore common
practice to estimate an unconstrained covariance matrix. We now show how
stochastic approximation methods can be used to impose the unit diagonal
constraint quickly and efficiently even in high dimensions.

Exact constraints in stochastic approximation are studied in~\cite{WangSpall1999}. For the problem at hand, a convenient implementation is to model the
constraint as a quadratic penalty term and iterate on
\begin{align*}
\bm{P}_{t}& =\text{diag}([\bm{V}_{t}]_{1,1}-1,...,[\bm{V}_{t}]_{p,p}-1) \\
\bm{V}_{t+1}& =\bm{V}_{t}+\frac{\alpha _{t}^{V}}{S}\left[ \sum_{i\in
T_{t}}(\bm{x}_{i}\bm{x}_{i}^{T}-\bm{V}_{t})-t\bm{P}_{t}\right] ,
\end{align*}%
where $T_{t}$ is a random draw with replacement
of $S$ elements (we use 20)
from the set $(1,...,n)$. The unconstrained covariance matrix provides a
good starting value. Notice that the penalty term is multiplied by the
iteration number $t$ which makes the constraint quite soft initially and
then progressively tighter. In our experience this delivers smooth and fast
convergence even in three-digit dimensions. The scalar sequence $%
\alpha _{t}^{V}$ is described in Appendix~\ref{B:stoch_approx}.

\subsection{$t$ copula} \label{SS: t copula}

We estimate a $t_{\nu}$ copula, that is, a $t$ copula with $\nu$ degrees of
freedom, as follows: (a)~Step 1 is the same as for a normal copula.
(b)~Fix the degrees of freedom parameter $\nu$. (c)~
Use the estimates in step 1 to construct the cumulative density of
each marginal and transform the original $\bm{y}$ into $\bm{x}$, where each element of
$\bm{x}$ is $t_{\nu}(0,1)$. (d)~Fit a multivariate $t$ distribution $t_{\nu}(\bm{0},\bm{V})$ to $\bm{x}$. This
implicitly defines a $t_{\nu}$ copula. Estimation is performed by iterating to
convergence on
\begin{equation}
\bm{V}_{t+1}^{-1}=\bm{V}_{t}^{-1}+0.5\sum_{i=1}^{n}\left[ \bm{V}_{t}-\frac{\nu+p}{\nu}\left( 1+%
\frac{x_{i}^{T}\bm{V}_{t}^{-1}x_{i}}{\nu}\right) ^{-1}\bm{x}_{i}\bm{x}_{i}^{T}\right] /n,
\end{equation}%
where a good starting value is provided by the method of moment estimate
\begin{equation*}
\widehat{\bm{V}}_{MM}=\frac{\nu-2}{\nu}\sum_{i=1}^{n}\bm{x}_{i}\bm{x}_{i}^{T}.
\end{equation*}

The likelihood of $\bm{y}$ can be computed using equation \eqref{eq: Py} where $%
f(\bm{x})$ is the multivariate $t$ density estimated in steps (b)--(d). The likelihood
can then be maximized with respect to the degrees of freedom parameter $\nu$
with standard optimization methods.

\section{Estimation of multivariate regression models with mixture of normal
errors by stochastic approximation\label{B:stoch_approx}}

The basic building block of our estimation methods is the mixture of normals
model, which makes it necessary to have fast and reliable methods of
estimating such models. The estimation of a mixture of normals is known to be
complicated by several factors: (i)~the likelihood is ill-defined at the
boundary of the parameter space, approaching infinity as the variance of any
component approaches zero; (ii)~it is necessary to estimate the number of
components in the mixture; (iii) the likelihood is typically multimodal;
(iv)~computing time can be high in large datasets or high dimensions.

The first problem is largely solved by placing weakly informative inverse
Wishart priors on the covariance matrices: $\bm{V}_{j}\sim IW(\bm{S},\nu)$.
Following \cite{fraley2005}, we set $\nu=1$ and $\bm{S}$ equal to the sample
variance of the data (or of the OLS residuals in the regression case).

We tackle the second problem by using the BIC information criterion, which
has been shown to perform reasonably well in this context (see \citet[][Chapter 6]{McLachlan2000} for a review).

The standard approach to estimating a mixture of normals is based on the EM
algorithm. For mixture of normals, the EM algorithm breaks a complex
maximization problem with no analytical solution into a sequence of simpler
maximizations with analytical solutions. Moreover, it requires no
user-defined tuning parameter other than a convergence criterion. Finally,
the EM algorithm is greedy, in the sense that it increases the likelihood at
each step. This greedy nature can be a weakness since mixture of
normal densities can be highly multimodal. Trying several starting
values is an effective strategy with a small number of parameters, but in higher
dimensions any random starting point is likely to be far from the global
mode due to the empty space phenomenon.

Several authors have proposed more principled strategies than random
starting values, by using the EM algorithm within split-and-merge strategies
\citep{figuereido2002}, genetic algorithms \citep{pernkopf2005},
or greedy search \citep{Verbeek2003}. We use stochastic approximation as a non-greedy alternative to the EM algorithm. \cite{Jordan94} and \cite{Yin2001}
document stochastic approximation
algorithms that largely outperform their EM counter-parts in mixture
problems, in both speed and quality of convergence.

We consider the $p$ dimensional multivariate model
\begin{equation}
{\bm{y}_{i}}=\bm{B}
{\bm{z}_{i}}+\bm{\epsilon} _{i},
\label{eq: MNregression}
\end{equation}%
where $\bm{z_i}$ is $k \times 1$ and
$\bm{\epsilon} _{i}$\ is $i.i.d$\ with a mixture of $m$\ normals density%
\begin{align*}
p(\bm{\epsilon} _{i}) &=\sum_{j=1}^{m}\pi _{j}\phi (\bm{\epsilon} _{i};\bm{\mu} _{j},
\bm{V}_{j}) \\
\pi _{j} &\geq 0,j=1,...,m,\ \sum_{j=1}^{m}\pi _{j}=1.
\end{align*}%
We leave the expected value of $\bm{\epsilon }$\ unconstrained and do not include
a constant in $\bm{z} $.

The stochastic approximation recursions given below are adapted from \cite{Yin2001} to include regression effects, a prior on the covariance
matrices, and batches of more than one observation. In these recursions the
gradient of the log-posterior is multiplied by the information matrix of the
\emph{complete} data log-likelihood as suggested by \cite{Titterington84}.

Let $\{\bm{y}_{i}, \bm{z}_{i}, i \in \bm{I}_t\}$\ be $S$\ observations drawn with replacement at
iteration $t,$\ with $\bm{I}_{t}$\ a vector of indices. \
Denote by $P(h|i,\theta _{k})$\ the probability that $\bm{y}_{i}$\ is generated
by component $h$\ given $\bm{z}_{i}$\ and $\theta _{k},$%
\begin{equation*}
P(h|t,\theta _{k})=\frac{\pi _{h}\phi (\bm{e}_{h,i};\bm{0},\bm{V}_{h,t})}{\sum_{j=1}^{m}\pi
_{j}\phi (\bm{e}_{j,i};\bm{0},\bm{V}_{j,k})},
\end{equation*}%
where $\bm{e}_{j,i}=\bm{y}_{i}-\bm{B}_{t}\bm{z}_{i}-\bm{\mu} _{j,k}$.

The parameters are updated using the following recursions:
\begin{align}
\bm{B}_{t+1}^{T} &=\bm{B}_{t}^{T}+\frac{\alpha _{t}^{B}}{S}\sum_{i\in
\bm{I}_{t}}\sum_{j=1}^{m}P(j|i,\theta _{t})\bm{V}_{z}^{-1}\bm{z}_{i}\bm{e}_{j,i}^{T}
\label{RecSAMN1} \\
\bm{V}_{j,t+1} &= \bm{V}_{j,t}+\frac{\alpha _{t}^{V}}{S}\sum_{i\in \bm{I}_{t}}P(j|i,\theta
_{t})(\bm{e}_{j,i}\bm{e}_{j,i}^{T}-\bm{V}_{j,t})+\frac{\alpha _{t}^{V}}{n}(\bm{S}/\nu -V_{j}),\text{
}j=1,...,m  \label{RecSAMN2} \\
\bm{\mu} _{j,t+1} &=\bm{\mu}_{j,t+1}+\frac{\alpha _{t}^{\mu }}{S}\sum_{i\in
\bm{I}_{t}}P(j|i,\theta _{t})\bm{e}_{i,t},\text{ }j=1,...,m  \label{RecSAMN3} \\
\pi _{j,t+1} &=\pi _{j,t+1}+\frac{\alpha _{t}^{\pi }}{S}\sum_{i\in
I_{t}}\biggl ( P(j|i,\theta _{t})-\pi _{j,t}\biggr ),\text{ }j=1,...,m.  \label{RecSAMN4}
\end{align}%
where $\bm{V}_{z}$\ is the sample variance of $\bm{z}$\ computed on all $n$\
observations. The recursions in \cite{Titterington84} divide the last terms by
$\pi _{j}$, but we find this to be less stable, particularly when
some components are redundant. Our approach follows that of \cite{Yin2001}.

The standard choice of sample size $S$\ is one. We prefer to use small
batches of 20 observations. This avoids excessively large updates in the
initial iterations, reduces the number of iterations needed for convergence
and allows a more efficient matrix implementation, while preserving
sufficient randomness to bounce off shallow local modes. We set the initial
values $\alpha _{0}^{B},\alpha _{0}^{\mu },\alpha _{0}^{V},\alpha _{0}^{\pi
} $\ to $(0.5,0.5,0.1,0.1).$ Their rate of decay is determined by the search
then converge formula of Darken et al. (2002).
\begin{equation*}
\alpha _{k}=\alpha _{0}\frac{1+\frac{c}{\alpha _{0}}\frac{k}{\tau }}{1+\frac{%
c}{\alpha _{0}}\frac{k}{\tau }+\tau (\frac{k}{\tau })^{2}},
\end{equation*}%
with $c=1$ and $\tau =100$.\ The rather small initial values increase
stability, while the large $\tau $\ extends the search phase and decreases
the chances of converging to a poor mode.\ As common in the SA literature,
the algorithm is run for a pre-set number of iterations ($1000$ in our case).

The linear coefficient matrix $\bm{B}$\ is initialized by OLS estimated on
de-meaned data. An inexpensive but effective way of initializing $(\bm{\mu}
_{1},...,\bm{\mu} _{m})$\ is to take points on the principal components of the
residuals from the OLS regression, as common in the estimation of
self-organizing maps \citep{Kohonen1990}, and equal probabilities and
covariances $\bm{V}_{j}=\bm{V}(\bm{e})/m,$ where
$\bm{e}_{i}=\bm{y}_{i}-\widehat{\bm{B}}_{OLS}\bm{z}_{i}.$

\section{Archimedian copulas} \label{A: copulas}
Archimedian copulas are of the form
$C(u_1, \dots, u_p)  = G^{-1} ( G(u_1) + \dots + G(u_p) )$ ,
where $G(\cdot)$ is called the generator and is strictly monotonic on $[0, \infty)$. The generators for the Clayton, Frank and Gumbel copulas are
$G(u) = (u^{-\theta} -1)/\theta $ with $\theta > 0 $,
$
G(u)  = -\log \biggl ( (\exp(-u\theta) -1)/(\exp(-\theta)-1) \biggr )
$ and
$G(u) = \biggl ( -\log(u) \biggr ) ^\theta$. See \cite[][Chapter~5]{McNeil05}.
We estimate the Archimedian copulas by maximum likelihood.

\section{Multivariate skew t distribution} \label{A: multivar skew t}
The multivariate skew $t$ distribution proposed by
\cite{Sahu03} can capture  both  skewness and kurtosis in data.
It is  of the form
$\bm{y}  = \bm{W\beta}  + \bm{Dz} + \bm{e} $ with $\bm{W\beta}$ , $\bm{V}$ and $\bm{D}$ the location, dispersion and skewness parameters.
$\bm{D}$ is a diagonal matrix with diagonal entries $\bm{\delta} = (\delta_1, \dots, \delta_p)$. The vector $\bm{z}$ has independent elements $z_i$ such that $z_i \sim t_\nu(0,1)I(z_i> 0 )$, that is truncated univariate
$t$ distributions with $\nu$ degrees of freedom. The disturbance $e \sim t_\nu (\bm{0},\bm{V})$, that is a multivariate $t$ distribution with $\nu$ degrees of freedom, zero mean and dispersion matrix $\bm{V}$. We estimate the skew-$t$ by a \MCMC{} simulation method as in \cite{Sahu03} and the parameters by their posterior means. We use the following prior:
$\bm{\beta} \sim N(\bm{0}, \bm{\Omega}_\beta)$,
$\bm{\delta} \sim N(\bm{0} , \bm{\Omega}_\delta)$, $\bm{V} \sim \text{Inverse Wishart} (m, \bm{\Omega}_V)$,
where $m$ is the degrees of freedom parameters and $\bm{\Omega}_V$ is the scale
matrix. The degrees of freedom parameter $\nu$ has a gamma prior with shape 1 and scale 20 and is truncated below at 2. The hyperparameters for the priors are
$\bm{\Omega}_\beta = \bm{\Omega}_\delta  =1000\bm{I}$ and
$\bm{\Omega}_V = (100/m)\bm{R}^{-1}$, where $m = 3$ and  $\bm{R}$ is a diagonal matrix with $i$th diagonal element the squared range of the corresponding element of the data.

\end{appendices}
\renewcommand{\baselinestretch}{1.0}
\newpage
\bigskip

\begin{table}[htbp]
\centering \fontsize{10pt}{22pt}\selectfont
\begin{tabular}{|r|r|r|r|r|r|r|r|r|}  \hline

    &    Clayton &      Frank &     Gumbel &         ST &         MN &       MAMN &         NC &        MNC \\ \hline

KL &      1.1472 &     0.9347 &     1.0183 &     2.6855 &     2.1963 &     1.5512 & 0.0161* & 0.0197* \\

   &      (0.0206) &     (0.0166) &     (0.0181) &     (0.0219) &     (0.0219 &     (0.0251) &     (0.0075) &     (0.0080) \\ \hline

L2 &      0.6245 &     0.3788 &     0.7226 &     1.2506 &     0.6278 &     1.6097 &     0.0585 &     0.0586 \\

   &       (0.0234) &     (0.0211) &     (0.0230) &     (0.0316) &     (0.0241) &     (0.0356) &     (0.0164) &     (0.0166) \\ \hline
\end{tabular}
\vspace{0.5cm}
		\caption{The data generating process is a normal copula. $p=5$ is the dimension and $n=500$ is the sample size. The table reports the median of the logarithm of the ratio of the loss for each estimator to the $t$ copula estimator. The standard errors are in brackets. A * means that we do not reject the null that the median is 0 at the 1\% level and a ** means that we do not reject at the 5\% level. }
\label{T:NC}
\end{table}

\begin{table}[htbp]
\centering \fontsize{10pt}{22pt}\selectfont
\begin{tabular}{|r|r|r|r|r|r|r|r|r|}  \hline
&       Frank &     Gumbel &         ST &         MN &       MAMN &         NC &         tC &        MNC \\ \hline

    KL &       2.1199 &     2.5710 &     2.7913 &     1.9638 &     1.9274 &     2.2835 &     2.0234 &     1.3965 \\

       &    (0.0474) &     (0.0492) &     (0.0479) &     (0.0522) &     (0.0503) &     (0.0499) &     (0.0496) &     (0.0446) \\ \hline

    L2 &       0.6772 &     0.7093 &     0.5846 &     0.2179 &     0.5526 &     0.6933 &     0.6017 &     0.4429 \\

       &     (0.0418) &     (0.0422) &     (0.0406) &     (0.0472) &     (0.0510) &     (0.0421) &     (0.0409) &     (0.0366) \\ \hline

\end{tabular}
\vspace{0.5cm}
\caption{The data is generated by a  Clayton copula with $\theta = 5$ and mixture of normals marginals. $p=5$ is the dimension and $n=500$ is the sample size. The table reports the median of the logarithm of the ratio of the loss for each estimator to the Clayton copula estimator.
The standard errors are in brackets. A * means that we do not reject the null that the median is 0 at the 1\% level and a ** means that we do not reject at the 5\% level.}
\label{T:Clayton}
\end{table}

\begin{table}[htbp]
\centering \fontsize{10pt}{22pt}\selectfont
\begin{tabular}{|r|r|r|r|r|} \hline
           &       ST &       MAMN &         tC &        MNC \\ \hline

        KL &      1.6874 &     0.1249 &     1.5458 &     1.4393 \\

           &    (0.0386) &   (0.0246) &   (0.0397) &   (0.0503) \\

               L2 &      1.2907 &  -0.0128** &     0.9502 &     0.4552 \\

           &    (0.0324) &   (0.0269) &   (0.0316) &   (0.0499) \\ \hline
\end{tabular}
\vspace{0.5cm}
		\caption{The data is generated by a scale mixture of normals. $p$ is the dimension and $n$ is the sample size. The table reports the median of the logarithm of the ratio of the loss for each estimator to the \mn{}
	 estimator. The standard errors are in brackets. A * means that we do not reject the null that the median is 0 at the 1\% level and a ** means that we do not reject at the 5\% level. }
\label{T:MN}
\end{table}

\begin{table}[htbp]
\centering \fontsize{10pt}{22pt}\selectfont
\begin{tabular}{|r|r|r|r|r|} \hline

           &          ST &         MN &         tC &        MNC \\ \hline

        KL &      1.0282 &     0.3351 &     0.5871 &     0.5963 \\

           &     (0.0115) &     (0.0100) &     (0.0119) &     (0.0113) \\ \hline

        L2 &      0.0975 &     0.4398 &     0.2540 &     0.2470 \\

           &     (0.0383) &     (0.0237) &     (0.0124) &     (0.0091) \\ \hline

\end{tabular}
\vspace{0.5cm}
\caption{The data is generated by a  mixture of normals with an additional uniform component. $p$ is the dimension and $n$ is the sample size. The table reports the median of the logarithm of the ratio of the loss for each estimator to the \mamn{} estimator. The standard errors are in brackets. A * means that we do not reject the null that the median is 0 at the 1\% level and a ** means that we do not reject at the 5\% level. }
 \label{T:MNU}
\end{table}

\begin{table}[htbp]
\centering \fontsize{12pt}{22pt}\selectfont
\begin{tabular}{|r|rrrrr|} \hline

           &         ST &         MN &       MAMN &        tC  &        MNC \\ \hline

        LPS &  -508.04 & -504.14 & -501.56 &  -504.39 &   -521.57 \\

      Rank &          4 &          2 &          1 &          3 &          5 \\

       NoC &          - &          2 &          2 &          - &          2 \\

       DoF &       3.83 &          - &          - &       4.20 &          - \\ \hline
\end{tabular}
\vspace{0.5cm}
		\caption{Results for the Fama and French three-factor model. LPS indicates the average log-predictive score from ten-fold cross validation. Rank is the ranking of the different models with 1 indicating the best model according to LPS and 5 the worst. NoC is the number of components found by the joint distribution estimation for each of the 10 cross-validation samples. It applies only to the mixture of normals, \mamn{} and the mixture of normals copula. DoF is the average degree of freedom over the 10 cross-validation samples. It  only applies to the skew t and t-copula models.}
\label{TableIndPF}
\end{table}

\begin{table}[htbp]
\centering \fontsize{12pt}{22pt}\selectfont
\begin{tabular}{|r|rrrrr|} \hline
           &         ST &         MN &       MAMN &        tC  &        MNC \\ \hline
        LPS &  -555.22 &  -480.70 &  -418.57 &  -478.07 &  -478.09 \\
      Rank &          5 &          4 &          1 &          2 &          3 \\
       NoC &          - &        2.4 &          2.4 &          - &        1.1 \\
       DoF &       9.21 &          - &          - &        $>30$ &          - \\ \hline
\end{tabular}
\vspace{0.5cm}
\caption{Realized volatility model. LPS is the average log-predictive score from ten-fold cross validation. Rank means the ranking of the different models with 1 indicating the best model according to LPS and 5 the worst. NoC means the average number of components found by the joint distribution estimation and applies only to the mixture of normals, \mamn{} and the mixture of normals copula. DoF is the average number of degrees of freedom, valid only for the skew t and t-copula models.}
\label{TableRV}
\end{table}
\begin{table}[htbp]
\centering \fontsize{12pt}{22pt}\selectfont
\begin{tabular}{|r|rrrrr|} \hline

           &         ST &         MN &       MAMN &        tC  &        MNC \\ \hline

       LPS &  -805.1310 &  -732.0859 &  -724.3464 &  -827.0198 &  -773.2771 \\

      Rank &          4 &          2 &          1 &          5 &          3 \\

       NoC &         -   &        5.3 &        5.3   &           - &        5.4 \\

       DoF &       5.14 &         -   &         -   &       8.10 &         -   \\ \hline

\end{tabular}
\vspace{0.5cm}
		\caption{Results for the gene expression data. LPS is the average log-predictive score from ten-fold cross validation. Rank means the ranking of the different models with 1 indicating the best model according to LPS and 5 the worst. NoC means the average number of components found by the estimation of the joint distribution and applies only to a mixture of normals, \mamn{} and mixture of normals copula. DoF is the average degrees of freedom, valid only for the skew t and t-copula model.}
\label{TableGene}
\end{table}

\end{document}